# Model-based Hardware Design for FPGAs using Folding Transformations based on Subcircuits

Konrad Möller*, Martin Kumm*, Charles-Frederic Müller†, Peter Zipf*
*Digital Technology Group, University of Kassel, Germany
†Volkswagen AG, Braunschweig, Germany
*Email: {konrad.moeller, kumm, zipf}@uni-kassel.de
†Email: charles-frederic.mueller@volkswagen.de

*Abstract*—We present a tool flow and results for a model-based hardware design for FPGAs from Simulink descriptions which nicely integrates into existing environments. While current commercial tools do not exploit some high-level optimizations, we investigate the promising approach of using reusable subcircuits for folding transformations to control embedded multiplier usage and to optimize logic block usage. We show that resource improvements of up to 70% compared to the original model are possible, but it is also shown that subcircuit selection is a critical task. While our tool flow provides good results already, the investigation and optimization of subcircuit selection is clearly identified as an additional keypoint to extend high-level control on low-level FPGA mapping properties.

## I. Introduction

The use of domain-specific modeling tools like Matlab/Simulink is a common way to describe (and test) data flow dominated applications, commonly denoted as model-based design. It has proven successful in automatic code generation which is the de-facto standard, e. g., in the automotive domain for many years. Up to 80% of the processor code in todays embedded control units is generated from Matlab/Simulink [1]. The increasing demand for processing high sample frequencies recently lead to performance requirements that exceed the capabilities of embedded CPUs. FPGAs provide a solution for this problem as they yield the required computational power. However, typical sample frequencies are still much lower than the FPGAs' system clock frequency. This opens the opportunity to reduce FPGA resources by computing parts of the design using time-multiplexing while sharing the computation modules.

A well known method to automatically transform a parallel data flow graph (DFG) into a sequential circuit is *folding* [2]. During the folding transformation, common operators like, e. g., multipliers are implemented only once and shared by using multiplexers and additional registers. These resources and the required controller introduce an overhead which has to be lower than the resources saved due to sharing to gain any benefit [3]. As the number of multiplexers directly scales with the number of inputs of shared operands, an overhead reduction could be obtained when operations are combined to larger subcircuits instead of equipping each single operation with multiplexers and registers. In principle, the same solution could be found with the right operator selection, scheduling and binding, but for this the right parameters have to be known.

The use of common subcircuits instead removes multiplexers and registers per construction. A subcircuit as defined in this work corresponds to a subgraph of the DFG. The more frequently a subcircuit occurs, the more resources can be saved due to sharing. However, the larger a common subcircuit is, the smaller is typically its frequency of occurrence. In addition, several independent common subcircuits may exist in the design which may even partly overlap, leading to a large design space. The task is thus related to a subgraph partitioning problem based on sets of isomorphic subgraphs. The target function on the other hand is based on implementation costs which are not directly related to individual subgraphs or to partition properties like size or number. Besides this, subcircuits can be used to reach a resource optimal point in the design space which meets the throughput requirements.

The idea behind our investigation is to apply well known high-level transformations to the Simulink description targeting resource reductions at the lowest register transfer/FPGA level. We present a tool flow which automatically utilizes the folding transformation to share arbitrary common subcircuits and show the benefits for this approach by a design space exploration of several benchmark circuits. The main contribution of this work is an extensive analysis of the results which were generated during this exploration. Besides this, we show that the results obtained with our tool flow are always better in terms of slices than the folding transformations of the Matlab HDL coder [4], which was taken as state-of-the-art reference in this work.

## II. Background

The *identification* of common subcircuits, also known as subcircuit recognition [5], subgraph enumeration [6]–[8] or clone detection [1], is a well known problem which appears in many different disciplines and is akin to the subgraph isomorphism problem which is known to be NP-hard [6]. Powerful methods have evolved in the last three decades. A good introduction into the topic can be found in [5]. However, the *beneficial use* of common subcircuits in synthesis is still not well understood. Common subcircuits have been used in high-level synthesis (HLS) tools targeting behavioral input languages like C [7]–[10]. A tool flow that enumerates subgraphs and re-uses them in the xPilot HLS tool was presented by Cong and Jiang [7]. They report FPGA resource reductions by about 20% on average. The sharing of single operations as well as common subcircuits (called composite operators or patterns) within the high-level synthesis (HLS) tool LegUp is analyzed in [10]. The sharing is limited to two operations with non-overlapping life times, i.e., one physical unit is used to compute two operations in the algorithm when no additional registers are required to store intermediate results. A greater benefit is reported when common subcircuits are used instead of single operations. They also analyzed the impact of the FPGA architecture and obtained area reductions from 7 to 12% by using subcircuit sharing.

Common subcircuits also have been used in folding to reduce the computation time of the folding transformation [11]. There, the overall system as well as a common subcircuit are folded separately which is called hierarchical folding. The result is identical to the folding of the complete circuit for single operations, but reduces the complexity for $M$ identical blocks from $\mathcal{O}(M^3)$ to $\mathcal{O}(M)$ (assuming the subcircuits are known in advance) [11]. A hierarchical synthesis methodology is also described in [9]. There, resource sharing is performed independently at each hierarchy level including controllers. To the best of our knowledge, common subcircuits were not used so far for resource reduction within the folding transformation.

## III. Folding Transformation

The folding transformation [2] is a systematic way to realize the time-multiplexed reuse of identical operations like e.g. additions and multiplications. In the implementation considered in this work, this can additionally be combinations of single operations to larger subcircuits which can be found more than once in the circuit. A common subcircuit or operation that is shared using folding is denoted *folding core* in the following. A circuit may consist of several common subcircuits which may be mutually exclusive, i.e., a subset of subcircuits may be selected as folding cores. Suitable common subcircuits used in the design space exploration are selected by the user in this work. This step could be automated by one of the subcircuit recognition methods discussed in [1], [5], [7], [8].

The folding procedure is illustrated by an example Simulink model of a discrete PI-controller as shown in Fig. 1(a). The possible folding cores could be a single product or a single addition, denoted as {*Prod*} and {*Add*} or the common subcircuit {*Prod,Add*} as highlighted in Fig. 1(a). In the next step, a scheduling is required to determine the time step in which each subcircuit will be executed. This is important to provide the right input data at the right time in the time-multiplexed circuit. In the given example this could be {*Prod_2,Add_1*} in the first time step and {*Prod_1,Add_2*} in the second time step for the folding with common subcircuits. The minimal number of required time steps which leads to a valid solution is called the *folding factor* $N$. In the best case it is equal to the number of identical subcircuits.

The scheduled processing times can be verified by the help of the *folding equation*, which determines the delay $D$ in number of clock cycles between two nodes $U$ and $V$ in the folded circuit:

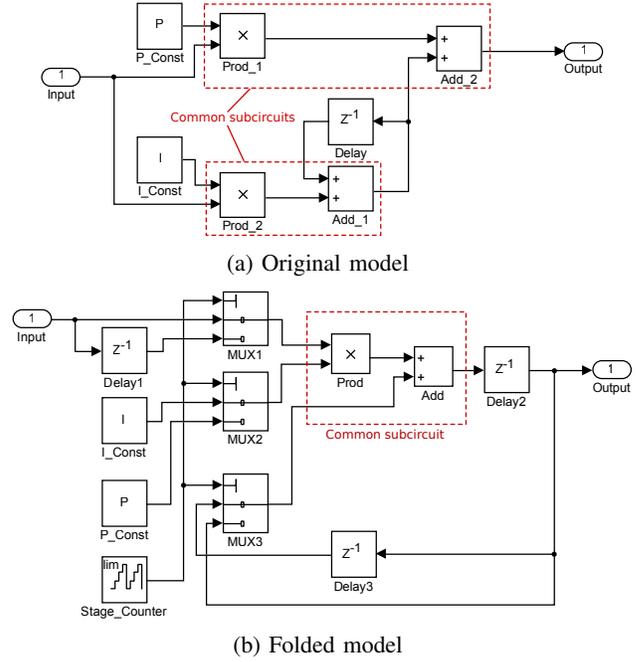

(a) Original model

(b) Folded model

Fig. 1: Example of a PI controller described in Simulink

$$D(U \xrightarrow{e} V) = Nw_e - P_u + v - u \geq 0 \quad (1)$$

where $w_e$ is the delay in number of clock cycles of the edge $e$ from nodes $U$ to $V$ in the original DFG, $P_u$ is the latency of $U$ and $u$ and $v$ are the scheduled execution times of $U$ und $V$, respectively.

Now, all the common subcircuits can be replaced by their corresponding folding cores which are equipped with multiplexers at their inputs. Each multiplexer input may require a number of additional registers ($D$) as given by the *folding equation*. Note that this may lead to a more complex wiring, but in our experiments this was not a limiting factor. The resulting folded PI controller circuit using the folding core {*Prod,Add*} is shown in Fig. 1(b).

To obtain a benefit from folding, the resulting circuit size $S_f$ of the $N$ times folded circuit has to be smaller than the size $S_o$ of the original circuit. Both circuits can be separated into the resources of the folding cores ($S_\text{folding core}$), non-folded parts ($S_\text{remain}$) and the overhead due to folding ($S_\text{overhead}$), leading to the condition:

$$S_f < S_o \quad (2)$$
$$S_\text{overhead} + S_\text{folding core} + S_\text{remain} < NS_\text{folding core} + S_\text{remain} \quad (3)$$
$$S_\text{overhead} < (N-1)S_\text{folding core} \quad (4)$$

Clearly, the overhead has to be smaller than the size of the saved folding cores. Thus, selecting a large folding factor and a large folding core should be worthwhile. But the capability to maximize both is limited by the structure of the original circuit, so a tradeoff has to be found.

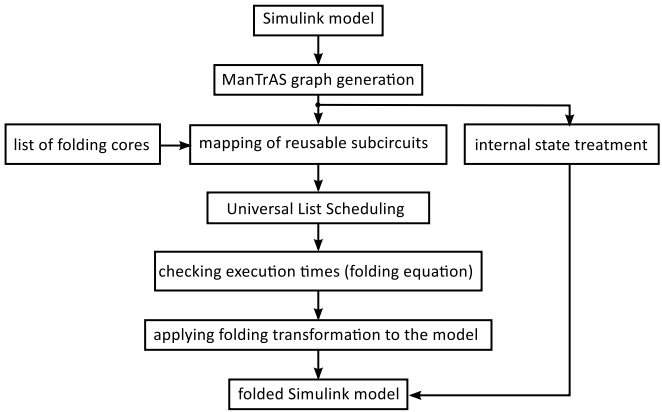

Fig. 2: Overview: Implemented folding transformation flow

## IV. Transformation Flow

In this section, the transformation flow of the implemented automatic folding transformation tool is presented. We perform structural transformations at the Simulink level to get the folded result. An overview is given in Fig. 2. The input of the transformation flow is the Simulink model of the design which is to be optimized. In the first step the model is transformed into an object oriented Matlab data structure by a tool called ManTrAS (**M**atlab **an**alysis and **Tr**ansformation **A**PI for **S**imulink) [12]. It provides an easy way to work with the data representing a Simulink model in order to analyze the model and to do the required transformations. Moreover a ManTrAS graph can be transformed back into a Simulink model. After the back-transformation everything which could be done with the original Simulink model can be applied to the folded one. This includes the simulation using the original test bench, HDL code generation using Mathworks' HDL Coder and validation of the folded model which is an essential advantage of this approach.

The next step is the mapping of the folding cores to the corresponding blocks in the ManTrAS graph. This is done by putting all blocks which belong to one folding core into a Simulink subsystem which can be identified by having a specific type of folding core. In contrast to other approaches we do not only consider one type of folding core but also a set of different non-overlapping folding cores for the folding transformation.

Since the former parallel design will be processed in a sequential time-multiplexed way after the transformation, the folding cores have to be assigned to time steps. This is currently done by a general list scheduling with support of resource constraints and multi-cycle operations [13]. The resource constraints are used to limit the available folding cores to one per clock cycle to force the schedule to be valid for the time-multiplexing. The latency $P_u$ of an operation is treated as a multi-cycle (blocking) operation in the scheduling. This guarantees that $v - u$ is always greater or equal than $P_u$ and, thus, the folding equation result is always positive or zero. The corresponding delays are inserted at the corresponding multiplexer input. A single control unit (a mod-$N$ counter) is inserted which controls the multiplexers to select the right input at the scheduled time step.

In addition to the original folding transformation, we support folding cores with internal states. For that, pipeline interleaving is applied to each folding core by simply replacing each register in the folding core by $N$ registers [2], [11].

## V. Experimental Setup

This section comprises an experimental design space exploration and evaluation of a benchmark set with different folding cores selected for the folding transformation. The results were automatically generated using the presented transformation flow. The evaluation is done with four commonly used applications in the domains control engineering and digital signal processing. The following applications have been chosen, implemented as functional Simulink models with a 32 bit fixed point precision and can be accessed online [14]:

A) 16 tap finite-impulse response (FIR) filter
B) Park-Clarke transformation (PCT)
C) Triple PID controller (TPID)
D) Infinite-impulse response (IIR) filter from [2]

While the FIR and IIR filters are well known, the Park-Clarke transformation is a combination of the alpha-beta transformation [15] and the direct-quadrature-zero transformation [16]. It is an important transformation in automotive controls and is used for the observing part of multiple-phase brushless DC motors. It consists of several sine look-up tables, additions/subtractions and multiplications.

The Triple PID controller is a design which implements three standard discrete PID controllers in parallel using the rectangular method for intergrals and differentials.

We distinguish between four implementation cases to compare the different folding strategies: The first case is the *original* unfolded design as reference. The second case is the *single operation folding*, which represents the folding strategy to share resource intensive single operations like, e. g., multiplications. The third case is the folding using *common subcircuits* as folding core, which was the main target of our exploration. The intention is mainly to show the benefits compared to the *single operation folding*. The HDL Coder resource sharing is taken as the last case, because it is a state-of-the-art commercial solution. The selected cores and their number can be found in TABLE I following the notation introduced in Sec. III. For example, in the FIR benchmark with folding factor $N = 5$ we reduce 5 cores each consisting of 2 delays, 2 products and 2 adders to 1 core and 2 cores consisting of 1 delay, 1 product and 1 adder to 1 core. This is denoted as 5{2delay,2prod,2add}, 2{delay,prod,add}. In some cases different folding cores are selected while the folding factor is the same, which results in different realizations for the same folding factor.

The folding transformation was performed for each selection using the presented flow, resulting in a folded Simulink model. This model was verified within the Simulink environment by a direct comparison to the input/output-behavior of the original unfolded model. After this step VHDL-Code was

9

TABLE I: Folding cores selected for the evaluation

| appl. | | folding cores | | |
|---|---|---|---|---|
| | N | cores | N | cores |
| FIR | 2 | 2{7delay,7prod,7add} | 14 | 14{prod,add,delay} |
| | 2 | 2{delay,prod} | 14 | 14{prod,delay} |
| | 3 | 3{4prod,4add, 4delay}, 2{2prod,2add,2delay} | 15 | 15{prod} |
| | 4 | 4{3prod,3add,3delay}, 3{prod,add} | 15 | 15{add}, 15{prod} |
| | 5 | 5{2delay,2prod,2add}, 2{delay,prod,add} | 15 | 15{prod,add} |
| | 7 | 7{2prod,2add,2delay} | | |
| PCT | 2 | 2{1conv,5sub, 3sin,3prod} | 6 | 6{sin}, 6{prod}, 2{prod}, 6{sub}, 4{sub} |
| | 2 | 2{sub,sin,prod}, 2{sub,sin,prod}, 2{sub,sin,prod} | 6 | 6{sin}, 6{prod}, 2{prod}, 3{sub}, 3{sub}, 2{sub}, 2{sub} |
| | 3 | 3{const,sub,sin,prod}, 3{sub,sin,prod} | 6 | 6{sub,sin,prod}, 2{prod}, 4{sub} |
| | 6 | 6{sin,prod} | 6 | 6{sin} |
| | 6 | 6{sin}, 6{prod}, 2{prod} | 6 | 6{sub,sin,prod} |
| TPID | 3 | 3{sub,prod,add} | 9 | 9{prod} |
| | 3 | 3{single PID} | 9 | 9{prod}, 9{add}, 6{sub} |
| | 3 | 3{P}, 3{I}, 3{D} | 9 | 9{prod,add} |
| | 6 | 6{sub,prod} | | |
| IIR | 2 | 2{prod,add,add} | 4 | 4{add}, 4{prod} |
| | 2 | 2{prod,add,prod,add} | 4 | 4{prod} |
| | 2 | 2{prod,add} | | |

generated using HDL Coder (v2.2) and functionally verified with ISim. Finally the VHDL code was synthesized for a Virtex 4 FPGA (xc4vlx200-10-ff1513) to get the required slice and DSP block count as well as timing information. The settings for DSP usage where set to *Auto*. The HDL Coder was used with different resource sharing factors which can be provided by the user to get a resource-latency tradeoff.

Special care has to be taken when constant values are included in the design. In such cases the constants may have an impact on the resulting design size, because different optimizations can be performed by the tools during the synthesis process. As we want to consider the general case we prevented the constants from being trimmed by replacing them by external inputs before synthesis. This means that the results in TABLE II and Fig. 3 and 4 represent an upper bound for the design sizes as for specific constants (like power-of-two values), less resources are possible.

## VI. RESULTS

A summary of the results can be found in TABLE II. The required resources of the unfolded (*original*) design are compared to the resources of the solution generated with the proposed folding flow using common subcircuits or single operation folding, which lead to the largest slice and DSP block reduction (*best fold.*) for the specific benchmark. The corresponding data points of these and the worse solutions can be found in Fig. 3. Moreover, the best result using common subcircuits (*comm. sub.*) is compared to the result of the corresponding single operation folding (*single op.*), as the investigation of their relation is the motivation of our work. A general observation is that a large amount of slice resources and DSP blocks can be saved compared to the original model and that the savings of common subcircuit folding typically surpass the savings of single operation folding.

TABLE II: Comparsion between original and best folded design and between common subcircuit and corresponding single operation folding for the example designs

| | FIR filter | | PCT | | Triple PID | | IIR filter | |
|---|---|---|---|---|---|---|---|---|
| | Slices | DSPs | Slices | DSPs | Slices | DSPs | Slices | DSPs |
| original | 663 | 64 | 12910 | 66 | 3492 | 96 | 135 | 16 |
| best fold. | 485 | 6 | 3938 | 19 | 1319 | 40 | 393 | 8 |
| *savings (%)* | *26* | *91* | *70* | *71* | *62* | *58* | *-191* | *75* |
| single op. | 1474 | 4 | 3938 | 19 | 1984 | 16 | 621 | 4 |
| comm. sub. | 485 | 6 | 4242 | 19 | 1731 | 16 | 428 | 4 |
| *savings (%)* | *67* | *-50* | *-7,7* | *0* | *13* | *0* | *31* | *0* |

In the following subsections specific observations for the different benchmarks and figures of the explored design space are provided (all numbers refer to TABLE II).

### A. 16 Tap FIR Filter

The results for the different FIR solutions can be found in Fig. 3 (a). For this very regular design the best solution can be achieved by a folding factor of 14 for the number of required slices and by a folding factor of 16 for the number of required DSP blocks. In order to achieve a slice reduction, the use of the largest common subcircuit (14{prod,add,delay}) is beneficial. The fact that the largest folding factor is leading to the smallest number of required DSP block is a general observation which holds for all analyzed benchmark circuits (A-D). For the FIR benchmark, there are many cases, especially with single operation folding, in which folding is leading to a much higher resource consumption compared to the unfolded design. This is the result of a large overhead compared to the saving which can be achieved by resource sharing for this rather small design. The best folded solution saves 26% of the required slices and 91% of the required DSP blocks compared to the original design. In the case of the FIR filter, common subcircuit folding is the only way to save slice resources. The corresponding single operation folding is not beneficial as the slice overhead is about 1000 slices, which exceeds the slice count of the original model.

### B. Park-Clarke Transformation

In the Park-Clarke Transformation example, the best solution can always be found with the largest folding factor. The analyzed cases have an almost identical slice consumption for identical folding factors as it can be seen in Fig. 3 (b). The HDL Coder was not able to perform any resource sharing independent of the given sharing factor. By application of our transformation flow on this rather large example we could achieve savings of about 70% for slice and DSP usage. The best solution for the single operation (6{sin},6{prod},2{prod},6{sub},4{sub}) and common subcircuit (6{sub,sin,prod},2{prod},4{sub}) folding have nearly the same slice consumption. This results from nearly identical folded solutions. In one case we define the common subcircuit (6{sub,sin,prod}) and in the other case each component of this common subcircuit is selected as a single operation (6{sin},6{prod},6{sub}). Based on a good scheduling for the single operation case, the architecture of the common subcircuit is reconstructed automatically during the transformation flow. The multiplexers in the single operation case thus have the same input at each port and can be replaced by a wire during synthesis.

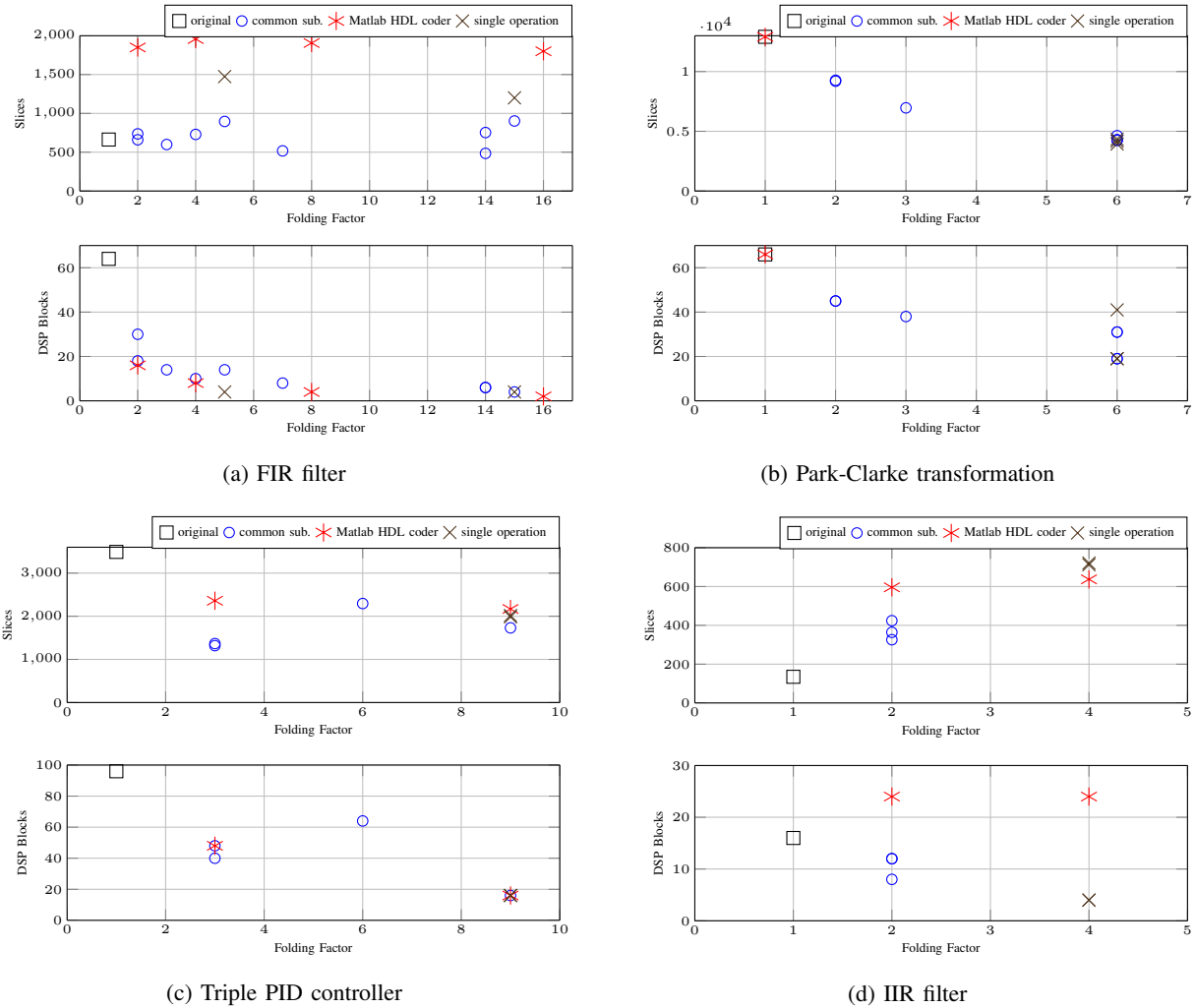

Fig. 3: Results for slice and DSP block usage for all benchmarks

## C. Triple PID Controller

The results for the Triple PID controller can be seen in Fig. 3 (c). The best case in terms of slice usage is not the case with the largest or nearly largest folding factor for this example. This can be explained by the fact that very large folding cores can be found in the best cases with only one input, leading to only one input multiplexer. The low overhead on the one side is further enhanced by the large resource saving on the other side, because of the large folding cores. In this case the folding core size was the defining element in the tradeoff between folding factor and folding core size (cf. (4)). The comparison of single operation (9{prod}) and the corresponding common subcircuit (9{prod,add}) folding in TABLE II shows again that it is beneficial to search for larger common subcircuits rather than folding around single operations.

## D. IIR Filter

The last example consists of only four multiplications and four additions and supplies only few folding alternatives. The results are plotted in Fig. 3 (d). The unfolded design is always the best solution in terms of slices, but choosing the multiplication and addition as folding core can significantly reduce the required DSP blocks. The HDL Coder resource sharing result is only able to save one out of four multipliers which leads to a larger slice overhead for time-multiplexing and no savings in DSP block consumption.

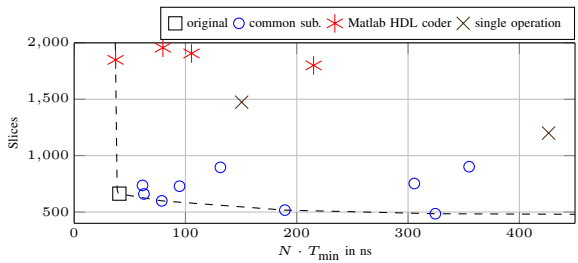

Fig. 4: Slice resource usage over $N \cdot T_{\min}$ of calculation for the FIR filter example and pareto front (dashed line)

### E. Resource Performance Tradeoff

With a set of different folding cores, leading to different folding factors and different latencies for data sample processing, our tool can deliver a tradeoff between resource consumption and latency of the processing of one data sample. The evaluated data points of the FIR filter example were used in Fig. 4 to show the number of required slices over the computation time $N \cdot T_{\min}$, where $T_{\min}$ denotes the minimal clock period obtained from the timing analysis. A designer with specific latency or area limitations could pick the best possible solution very easily. For a sample period requirement the best solution in terms of slice usage can be found as the point at the very bottom which is left of the latency limit and for a slice limitation the best solution in terms of latency is the leftmost point which is lower than the slice limit.

### F. Summary

The experimental results show that it is beneficial in terms of slice reduction to use common subcircuits instead of single operations as folding cores. Besides the slice reduction the selection of common subcircuits always leads to the best results if low latency is required. The largest DSP block reduction can be achieved by selecting the maximum folding factor in all cases, which can lead to a large slice overhead on the other hand. The relation of overhead and reduction by folding factor and/or folding core size (4) could be seen in some cases, but further investigations have to be done. The subcircuit selection itself is a critical task and it has a significant impact on the resulting design. However, the design space exploration which is possible with the presented transformation flow can help to find the best solution in order to fulfill the application constraints. The design space exploration considered the folding factor as well as the folding core size. A change in the degree of sharing, i.e., varying the amount of available folding cores was not analyzed, but should be considered in future work, as it delivers an additional optimization possibility.

## VII. CONCLUSION

We presented an automated high-level transformation for Simulink models targeting optimized FPGA implementations. The resulting models can be used by standard tools to generate VHDL code. We showed results only for four models but it becomes evident that very large improvements can appear, up to 70% in our examples. Using the tool flow, a design space exploration concerning the folding factor and the folding core inputs could be established and first results were analyzed in this paper. The results clearly show the benefits of our approach but also, that the selection of folding cores is a key factor for the improvement of FPGA resource requirements for any specific model: while the usage of embedded multipliers can be directly controlled, logic block requirements depend largely on the selected subcircuits.

Currently, different core combinations are selected by the user, but the results indicate that there is a non-trivial dependency between the model structure, the folding factor and the subcircuits and their possible combinations. An algorithm for subcircuit selection and combination is obviously necessary to fully automate the process of defining reasonable folding cores as input to our tool flow. The development of such an heuristic and the definition of according selection criteria are therefore the main targets of our future work.

## REFERENCES

[1] F. Deissenboeck, B. Hummel, E. Jurgens, B. Schatz, S. Wagner, J. F. Girard, and S. Teuchert, "Clone detection in automotive model-based development," in *Software Engineering, 2008. ICSE '08. ACM/IEEE 30th International Conference on*, 2008, pp. 603–612.

[2] K. K. Parhi, *VLSI Digital Signal Processing Systems: Design and Implementation*. John Wiley & Sons, 1999.

[3] W. Sun, M. J. Wirthlin, and S. Neuendorffer, "FPGA Pipeline Synthesis Design Exploration Using Module Selection and Resource Sharing," *Computer-Aided Design of Integrated Circuits and Systems, IEEE Transactions on*, vol. 26, no. 2, pp. 254–265, 2007.

[4] G. Venkataramani, K. Kintali, S. Prakash, and S. van Beek, "Model-based hardware design," in *Computer-Aided Design (ICCAD), 2013 IEEE/ACM International Conference on*. IEEE, 2013, pp. 69–73.

[5] N. Rubanov, "A High-Performance Subcircuit Recognition Method Based on the Nonlinear Graph Optimization," *Computer-Aided Design of Integrated Circuits and Systems, IEEE Transactions on*, vol. 25, no. 11, pp. 2353–2363, 2006.

[6] J. L. White, M. J. Chung, A. S. Wojcik, and T. E. Doom, "Efficient algorithms for subcircuit enumeration and classification for the module identification problem," in *International Conference on Computer Design (ICCD)*. IEEE, 2001, pp. 519–522.

[7] J. Cong and W. Jiang, "Pattern-based behavior synthesis for FPGA resource reduction," in *Proceedings of the 16th international ACM/SIGDA symposium on FPGAs*, New York, USA, Feb. 2008, pp. 107–116.

[8] J. Cong, B. Liu, S. Neuendorffer, J. Noguera, K. Vissers, and Z. Zhang, "High-Level Synthesis for FPGAs: From Prototyping to Deployment," *Computer-Aided Design of Integrated Circuits and Systems, IEEE Transactions on*, vol. 30, no. 4, pp. 473–491, 2011.

[9] O. Bringmann and W. Rosenstiel, "Resource sharing in hierarchical synthesis," in *Computer-Aided Design, IEEE International Conference on*. IEEE, 1997, pp. 318–325.

[10] S. Hadjis, A. Canis, J. H. Anderson, J. Choi, K. Nam, S. Brown, and T. Czajkowski, "Impact of FPGA architecture on resource sharing in high-level synthesis," pp. 111–114, 2012.

[11] K. K. Parhi, "Hierarchical Folding and Synthesis of Iterative Data Flow Graphs," *IEEE Transactions on Circuits and Systems II: Express Briefs*, vol. 60, no. 9, pp. 597–601, 2013.

[12] C. Kolassa, D. Dieckow, M. Hirsch, U. Creutzburg, C. Siemers, and B. Rumpe, "Objektorientierte Graphendarstellung von Simulink-Modellen zur einfachen Analyse und Transformation," *Tagungsband AALE 2013, 10. Fachkonferenz*, pp. 277–286, 2013.

[13] G. De Micheli, *Synthesis and Optimization Of Digital Circuits*. McGraw-Hill, 2003.

[14] K. Möller, M. Kumm, and C.-F. Müller, "Simulink Benchmarks and Results," http://www.uni-kassel.de/go/simulink_benchmarks, 2015.

[15] E. Clarke, *Circuit Analysis of AC Power Systems*. Wiley & Sons, 1943.

[16] R. H. Park, "Two Reaction Theory of Synchronous Machines Generalized Method of Analysis-Part I," *AIEE Transactions*, vol. 48, pp. 716–727, 1929.